\documentclass[a4paper]{article}

\usepackage{CJK}
\usepackage[paperwidth=185mm,paperheight=230mm,textheight=190mm,textwidth=145mm,left=20mm,right=20mm, top=25mm, bottom=20mm]{geometry}
\usepackage[CJKbookmarks, colorlinks, bookmarksnumbered=true,pdfstartview=FitH,linkcolor=blue,citecolor=green]{hyperref}
\usepackage{amsmath,amssymb}
\usepackage{amsthm}
\usepackage{calc}
\usepackage{graphicx}
\usepackage{supertabular}
\usepackage{longtable}
\usepackage{float}
\usepackage{color}
\usepackage{enumerate}
\usepackage{colortbl,booktabs}
\usepackage{multirow}
\newtheorem{theorem}{Theorem}

\usepackage{natbib}
\usepackage{url}
\usepackage{bm}
\usepackage{enumitem}
\usepackage{authblk}

\def\wh{\widehat}
\def\wt{\widetilde}

\begin{document}

\title{ Nonlinear Functional Principal Component Analysis Using Neural Networks }
\author[a]{{\fontsize{12pt}{18pt}\selectfont Rou Zhong}}
\author[b]{{\fontsize{12pt}{0.5em}\selectfont Chunming Zhang}}
\author[a]{{\fontsize{12pt}{0.5em}\selectfont Jingxiao Zhang} \thanks{zhjxiaoruc@163.com}}
\affil[a]{{\emph\fontsize{12pt}{0.5em}\selectfont Center for Applied Statistics, School of Statistics, Renmin University of China}}
\affil[b]{{\emph\fontsize{12pt}{0.5em}\selectfont Department of Statistics, University of Wisconsin-Madison}}
\date{}
\maketitle

\begin{abstract}

Functional principal component analysis (FPCA) is an important technique for dimension reduction in functional data analysis (FDA). Classical FPCA method is based on the Karhunen-Lo\`{e}ve expansion, which assumes a linear structure of the observed functional data. However, the assumption may not always be satisfied, and the FPCA method can become inefficient when the data deviates from the linear assumption. In this paper, we propose a novel FPCA method that is suitable for data with a nonlinear structure by neural network approach. We construct networks that can be applied to functional data and explore the corresponding universal approximation property. The main use of our proposed nonlinear FPCA method is curve reconstruction. We conduct a simulation study to evaluate the performance of our method. The proposed method is also applied to two real-world data sets to further demonstrate its superiority.


\end{abstract}

\section{Introduction}

Functional data analysis (FDA) has become widely concerned, with the rapid development of data collection technology. There are many monographs that provide a detailed introduction of FDA, such as \citet{ramsay2005functional}, \citet{ferraty2006nonparametric} and \citet{horvath2012inference}. Functional principal component analysis (FPCA), as a dimension reduction technique, plays a greatly important role in FDA, since functional data is a type of data with infinite dimensional. However, traditional FPCA is merely a linear approach, and the linear assumption can limit the effectiveness of dimensional reduction. In this paper, we aim to develop a nonlinear FPCA method based on the use of neural networks.

Recently, neural network approach draws more and more attention in the field of FDA and shows strong potential. \citet{thind2023deep} proposed a functional neural network to handle nonlinear regression model with functional covariates and scalar response. Further, \citet{rao2023non} introduced a continuous layer in the construction of functional neural networks, so that the functional nature of the data can be maintained as long as possible. The nonlinear function-on-scalar regression model has been considered by \citet{wu2022neural} with the use of neural networks. Moreover, neural network method is also employed in the classification problem of functional data, such as \citet{thind2020neural} and \citet{wang2022deep}. The above works all focus on supervised learning, as the regression or classification issues for functional data are considered, where label variables are defined. Furthermore, unsupervised learning problems for functional data have also been studied by neural networks. \citet{wang2021estimation} discussed the mean function estimation for functional data using neural networks. Multi-dimensional functional data is taken into account by \citet{wang2022robust}, and a robust location function estimator is proposed via neural networks. \citet{sarkar2022covnet} concentrated on covariance estimation for multi-dimensional functional data, and three types of covariance networks are defined correspondingly. Though neural networks have gained extensive interest in FDA, there are only very few studies working on nonlinear dimensional reduction for functional data through neural networks. \citet{wang2022functional} presented a functional nonlinear learning method, which is a representation learning approach for multivariate functional data and can be applied to curve reconstruction and classification. However, their method is developed based on recurrent neural network (RNN), thus only discrete values of the data are used in the neural network. Therefore, a nonlinear dimension reduction method by neural networks that treats the continuously observed data from a functional perceptive is needed.

FPCA is a crucial dimension reduction tool in FDA. There has been a great many works contributing to the development of FPCA in various aspects. These include, but are not limited to the study of principal component analysis for sparsely observed functional data \citep{yao2005functional, hall2006properties, li2010uniform}. Robust FPCA approaches were introduced in \citet{locantore1999robust, gervini2008robust} and \citet{zhong2022robust}. Moreover, \citet{chiou2014multivariate} and \citet{happ2018multivariate} discussed principal component analysis methods for more complex functional data, such as multivariate functional data and multi-dimensional functional data. For nonlinear FPCA, \citet{song2021nonlinear} generalized kernel principal component analysis to accommodate functional data. Currently, research on nonlinear FPCA is not sufficient enough. Nevertheless, the consideration of nonlinear structure of functional data can be beneficial, since more parsimonious representation can be obtained.

To this end, we propose a new nonlinear functional principal component analysis method by neural networks, which can be simply denoted as nFunNN. In specific, we borrow the idea of the autoassociative neural networks in \citep{kramer1991nonlinear} for the construction of our networks, to realize dimension reduction and curve reconstruction. \citet{kramer1991nonlinear} achieved the purpose of dimension reduction for multivariate data through an internal ``bottleneck" layer. However, the extension to functional data is nontrivial due to the infinite dimensional nature of functional data, which adds the complexity of the neural networks and increases the difficulty in the optimization. For our proposed neural network, both input and output are functions. To the best of our knowledge, though neural networks with functional input have been studied in the existing works, networks with both functional input and functional output have not been taken into account yet, and the consideration of which can be more complicated. B-spline basis functions are employed in our computation and backpropagation algorithm is applied. The simulation study and real data application show the superiority of the nFunNN method under various nonlinear settings. Moreover, we also establish the universal approximation property of our proposed nonlinear model.

The contributions of this paper can be summarized as follows. First, our work is the first attempt to the generalization of the autoassociative neural networks to functional data settings, which is not straightforward. The use of neural networks provides new framework of nonlinear dimension reduction for functional data. Second, the universal approximation property of the proposed model is discussed, which brings theoretical guarantees to our method. Third, we present an innovative algorithm for the computation in practice and develop a python package, called \texttt{nFunNN}, for implementation.

The organization of this paper is as follows. In Section \ref{SecMethod}, we first give an explanation of nonlinear FPCA, and then introduce a functional autoassociative neural network to complete nonlinear principal component analysis for functional data. We also discuss the practical implementation of our method. In Section \ref{SecSim}, we display the simulation results in our numerical study. The evaluation of our method by real-world data is provided in Section \ref{SecReal}. In Section \ref{SecDis}, we conclude this paper with some discussions.

\section{Methodology}\label{SecMethod}

\subsection{Nonlinear FPCA}

Let $X(t)$ be a smooth random function in $L^2(\mathcal{T})$, where $\mathcal{T}$ is a bounded and closed interval. In this paper, $\mathcal{T}$ is set as $[0, 1]$ if there is no specific explanation. Let $\mu(t)$ and $\Sigma(s, t)$ denote the mean function and covariance function of $X(t)$, respectively. For linear FPCA, according to the Karhunen-Lo\`{e}ve Theorem, $X(t)$ admits the following expansion
\begin{align}
X(t) = \mu(t) + \sum_{k = 1}^{\infty} \xi_k \phi_k(t), \nonumber
\end{align}
where $\xi_k$ is the $k$-th functional principal component score, $\phi_k(t)$ is the $k$-th eigenfunction of $\Sigma(s, t)$ and satisfies $\int_{\mathcal{T}} \phi_k(t)^2 dt = 1$ and $\int_{\mathcal{T}} \phi_k(t) \phi_l(t) dt = 0$ for $l \neq k$. Moreover, the functional principal component scores are uncorrelated random variables with mean zero and $E \xi_k^2 = \lambda_k$, where $\lambda_k$ is the $k$-th eigenvalue of $\Sigma(s, t)$. In practice, as only the first several functional principal component scores dominate the variation, a truncated expansion
\begin{align}
X(t) = \mu(t) + \sum_{k = 1}^{K} \xi_k \phi_k(t) \label{KLexptrun}
\end{align}
is often applied, where $K$ is the number of functional principal components that used.
Furthermore, we also have that
\begin{align}
\xi_k = \int_{\mathcal{T}} \{X(t) - \mu(t)\} \phi_k(t) dt. \label{scorelinear}
\end{align}
It is obvious that $X(t)$ is mapped into a lower dimensional space through a linear transformation. However, with the constraint of linear map, the nonlinear structure is ignored, which may lower the efficiency of dimension reduction.

For nonlinear FPCA, we extend the linear map in (\ref{scorelinear}) to arbitrary nonlinear map. That is
\begin{align}
\xi_k = G_k(X), \label{G_kX}
\end{align}
where $G_k$ is a nonlinear function that maps function in square integrable space $L^2(\mathcal{T})$ to scalar in $\mathbb{R}$. Similarly, (\ref{KLexptrun}) can be generalized into nonlinear version as
\begin{align}
X(t) = H(\bm{\xi})(t), \nonumber
\end{align}
where $\bm{\xi} = (\xi_1, \ldots, \xi_K)^\top$ and $H$ is a nonlinear function that maps from a vector space to a square integrable space. The scores obtained from nonlinear FPCA also contain the main information of $X(t)$, but the dimension of the feature space can be lower, as the nonlinear structure is taken into account in the process of dimension reduction. To estimate the nonlinear functional principal component scores, the nonlinear functions $G_k$ and $H$ have to be learnt, and a neural network method is employed.

\subsection{Neural Networks for Nonlinear FPCA}\label{SecNN}

\begin{figure}[H]
  \centering
  \includegraphics[width=0.8\textwidth]{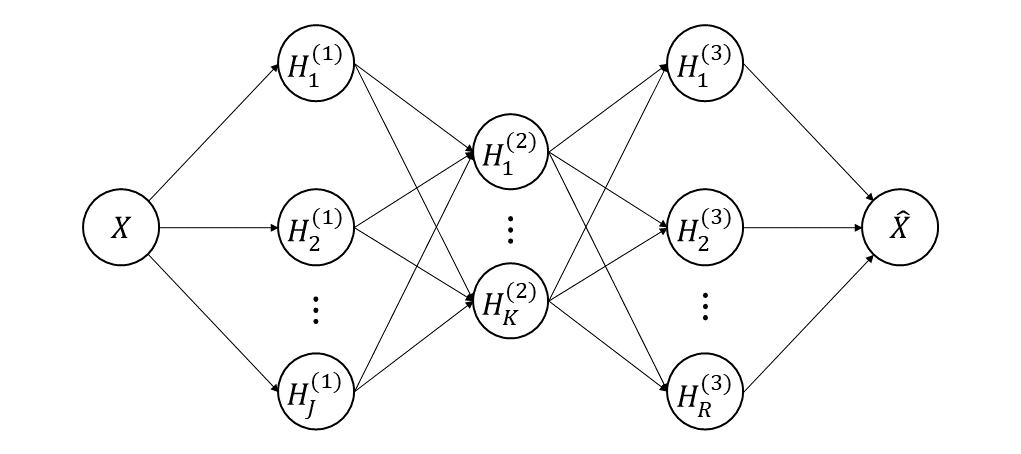}\\
  \caption{The proposed functional autoassociative neural network for nonlinear FPCA.}
  \label{FigNN}
\end{figure}

In this section, we construct a functional autoassociative neural network for nonlinear FPCA. The structure of the proposed neural network is shown in Figure \ref{FigNN}. The output is the reconstruction of the input data, thus both input and output are functions in our network, which brings challenge to the optimization of the neural network. Furthermore, dimension reduction is realized by the second hidden layer, which can also be called ``bottleneck" layer as in \citep{kramer1991nonlinear}. More details about the computation of the proposed functional autoassociative neural network are given below.

To be specific, the left two hidden layers are designed to learn the nonlinear functions $G_k$'s that map the functional input to the scores, which can be viewed as a dimension reduction process.
For the $j$th neuron in the first hidden layer, we define that
\begin{align}
H_j^{(1)} = \sigma \Big \{ b_j + \int_{\mathcal{T}} \beta_j(t) X(t) dt \Big \}, \ j = 1, \ldots, J, \nonumber
\end{align}
where $b_j \in \mathbb{R}$ is the intercept, $\beta_j(t) \in L^2(\mathcal{T})$ is the weight function, $\sigma(\cdot)$ is a nonlinear activation function, and $J$ is the number of neurons in the first hidden layer. This is a natural generalization of the first two layers of a multilayer perceptron to adapt to the functional input. As $H_j^{(1)} \in \mathbb{R}$, computation of the score in the second hidden layer can be promoted naturally. The $k$th neuron in the second hidden layer is
\begin{align}
H_k^{(2)} = \sum_{j = 1}^J w_{jk} H_j^{(1)}, \ k = 1, \ldots, K, \nonumber
\end{align}
where $w_{jk} \in \mathbb{R}$ is the weight. Moreover, let $\mathcal{S}(\sigma, L^2(\mathcal{T}))$ be the set of functions from $L^2(\mathcal{T})$ to $\mathbb{R}$ of the form
\begin{align}
X \mapsto \sum_{j = 1}^J w_{jk} \sigma \Big \{ b_j + \int_{\mathcal{T}} \beta_j(t) X(t) dt \Big \}. \nonumber
\end{align}
According to Corollary 5.1.2 in \citep{stinchcombe1999neural} and the proof in Section 6.1.2 of \citep{rossi2006theoretical}, $\mathcal{S}(\sigma, L^2(\mathcal{T}))$ has the universal approximation property for $L^2(\mathcal{T})$. That means any nonlinear function from $L^2(\mathcal{T})$ to $\mathbb{R}$ can be approximated up to arbitrary degree of precision by functions in $\mathcal{S}(\sigma, L^2(\mathcal{T}))$.

The right two layers in Figure \ref{FigNN}, which correspond to the estimation of the nonlinear function $H$, are used to reconstruct $X(t)$ by the low-dimensional scores $H_k^{(2)}$ in the second hidden layer. The procedure can be more challenging, since we have to get functional output from scalars in the second hidden layer. To this end, the $r$th neuron in the third layer is defined as
\begin{align}
H_r^{(3)} (t) = \sigma \Big \{ a_r(t) + \sum_{k = 1}^K \gamma_{kr} (t) H_k^{(2)} \Big \}, \ t \in \mathcal{T}, \ r = 1, \ldots, R, \nonumber
\end{align}
where $a_r (t) \in L^2(\mathcal{T})$ is the intercept function, $\gamma_{kr}(t) \in L^2(\mathcal{T})$ is the weight function, and $R$ is the number of the neurons in the third hidden layer. It can be observed that each neuron in the third hidden layer is a function. Then, $X(t)$ is reconstructed by
\begin{align}
\wh{X}(t) = \sum_{r = 1}^R u_r H_r^{(3)} (t), \ t \in \mathcal{T}, \label{NNlayerFour}
\end{align}
where $u_r \in \mathbb{R}$ is the weight. The whole network is trained by minimizing the following reconstruction error
\begin{align}
RE = \int_{\mathcal{T}} \{ X(t) - \wh{X}(t) \}^2 dt. \nonumber
\end{align}
As functional data is involved in the proposed network, some of the parameters need to be estimated are functions, which makes the optimization of the network nontrivial. In Section \ref{SecPraImp}, we introduce the optimization algorithm for practical implementation.

Note that $H_k^{(2)}$ can be viewed as the estimation of $\xi_k$ in (\ref{G_kX}). Therefore, the dimension of $X$ is reduced to $K$ through the functional autoassociative neural network. We can use the low-dimensional vector to complete further inference, such as curve reconstruction, regression and clustering. In this paper, we mainly focus on the curve reconstruction by the low-dimensional representation.

\subsection{The Transformed Network for Practical Implementation}\label{SecPraImp}

As discussed in Section \ref{SecNN}, it can be hard to optimize the proposed functional autoassociative neural network in Figure \ref{FigNN}, since many parameters appear as a function. Here, we employ the B-spline basis functions to transform the estimation of functions to the estimation of their coefficients. Let $\textbf{B}_{L} = \{B_l(t), l = 1, \ldots, L\}$ be a set of B-spline basis functions with degree $d$, where $L$ is the number of basis functions. Then, we have
\begin{align}
\beta_j(t) = \sum_{l = 1}^L c_{jl} B_l(t), \ a_r(t) = \sum_{l = 1}^L \alpha_{rl} B_l(t), \ \gamma_{kr}(t) = \sum_{l = 1}^L v_{krl} B_l(t), \nonumber
\end{align}
for $j = 1, \ldots, J$, $k = 1, \ldots, K$, and $r = 1, \ldots, R$, where $c_{jl}$, $\alpha_{rl}$ and $v_{krl}$ are the basis expansion coefficients of $\beta_j(t)$, $a_r(t)$ and $\gamma_{kr}(t)$, respectively.

With the use of B-spline basis functions, the computation of the first two layers for the proposed functional autoassociative neural network turns out to be
\begin{align}
H_k^{(2)} = \sum_{j = 1}^J w_{jk} \sigma \Big \{ b_j + \sum_{l = 1}^L c_{jl} \int_{\mathcal{T}} X(t) B_l(t) dt \Big \} \triangleq \sum_{j = 1}^J w_{jk} \sigma \Big ( b_j + \sum_{l = 1}^L c_{jl} \wt{X}_l \Big ), \label{NNtranlayertwo}
\end{align}
where $\wt{X}_l = \int_{\mathcal{T}} X(t) B_l(t) dt$. In practice, $\wt{X}_l$ is calculated through the B-spline expansion of $X$, that is $\wt{X}_l = \sum_{h = 1}^L x_h \{ \int_{\mathcal{T}} B_h(t) B_l(t) dt \}$, where $x_h$ is the basis expansion coefficient of $X(t)$. Then, we have
\begin{align}
H_k^{(2)} = \sum_{j = 1}^J w_{jk} \sigma \Big [ b_j + \sum_{h = 1}^L \Big \{ \sum_{l = 1}^L c_{jl} \int_{\mathcal{T}} B_h(t) B_l(t) dt \Big \} x_h \Big ] \triangleq \sum_{j = 1}^J w_{jk} \sigma \Big ( b_j + \sum_{h = 1}^L d_{jh} x_h \Big ), \nonumber
\end{align}
where $d_{jh} = \sum_{l = 1}^L c_{jl} \int_{\mathcal{T}} B_h(t) B_l(t) dt$. The following theorem discusses the universal approximation property of the first two layers based on B-spline basis functions. The proof is provided in the Appendix.

\begin{theorem}\label{TheUA}
Let $\sigma$ be a continuous non polynomial function from $\mathbb{R}$ to $\mathbb{R}$, and $\mathcal{S} (\sigma, \textbf{B}_L)$ be the set of functions from $L^2(\mathcal{T})$ to $\mathbb{R}$ of the form
\begin{align}
X \mapsto \sum_{j = 1}^J w_{j0} \sigma \Big ( b_j + \sum_{h = 1}^{L} d_{jh} x_h \Big ), \nonumber
\end{align}
where $x_h$ is the $h$th coordinate of $X$ on the basis $\textbf{B}_L$, $J \in \mathbb{N}^{\ast}$, $w_{j0} \in \mathbb{R}$, $b_j \in \mathbb{R}$ and $d_{jh} \in \mathbb{R}$. Then, $\mathcal{S} (\sigma, \textbf{B}_L)$ has the universal approximation property. That is for any compact subset $\mathbb{K}$ of $L^2(\mathcal{T})$, for any $F$ from $\mathbb{K}$ to $\mathbb{R}$ and for any $\epsilon > 0$, there exists $G \in \mathcal{S} (\sigma, \textbf{B}_L)$ such that for all $X \in \mathbb{K}$, $|G(X) - F(X)| < \epsilon$.
\end{theorem}

\begin{figure}
  \centering
  \includegraphics[width=0.4\textwidth]{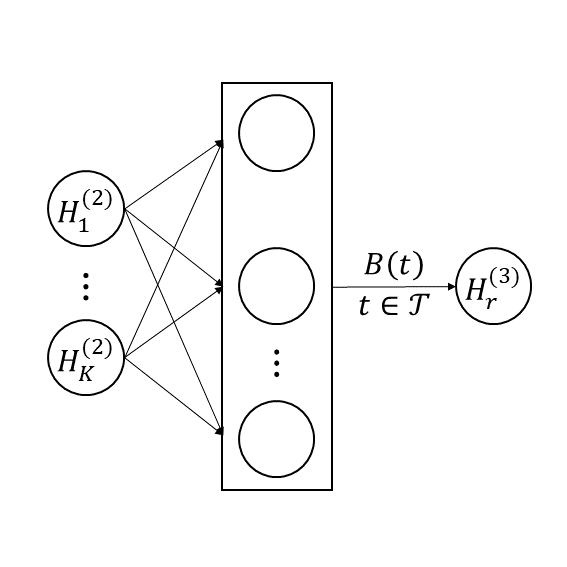}\\
  \caption{The network structure corresponding to the computation of (\ref{NNtranlayerthree}).}
  \label{FigNNsub}
\end{figure}

For the computation of the third hidden layer of the proposed functional autoassociative neural network, we have
\begin{align}
a_r(t) + \sum_{k = 1}^K \gamma_{kr} (t) H_k^{(2)} &= \sum_{l = 1}^L \alpha_{rl} B_l(t) + \sum_{k = 1}^K \sum_{l = 1}^L v_{krl} B_l(t) H_k^{(2)} = A_r^\top B(t) + \sum_{k = 1}^K V_{kr}^\top B(t) H_k^{(2)} \nonumber \\
&= \Big ( A_r + \sum_{k = 1}^K V_{kr} H_k^{(2)} \Big )^\top B(t), \nonumber
\end{align}
where $A_r = (\alpha_{r1}, \ldots, \alpha_{rL})^\top$, $V_{kr} = (v_{kr1}, \ldots, v_{krL})^\top$, and $B(t) = (B_1(t), \ldots, B_L(t))^\top$. For the term $A_r + \sum_{k = 1}^K V_{kr} H_k^{(2)}$ above, it can be viewed as weighted sums of $H_1^{(2)}, \ldots, H_K^{(2)}$. Thus, the computation from the second hidden layer to the third hidden layer of the functional autoassociative neural network can be transformed accordingly, that is
\begin{align}
H_r^{(3)}(t) = \sigma \Big \{ \Big ( A_r + \sum_{k = 1}^K V_{kr} H_k^{(2)} \Big )^\top B(t) \Big \}, t \in \mathcal{T}, \label{NNtranlayerthree}
\end{align}
the structure of which is shown in Figure \ref{FigNNsub}. The middle layer in Figure \ref{FigNNsub} represents the elements of the $L$-dimensional vector $A_r + \sum_{k = 1}^K V_{kr} H_k^{(2)}$. Furthermore, the reconstruction of $X(t)$ can be obtained by (\ref{NNlayerFour}) as discussed in Section \ref{SecNN}.

\begin{figure}
  \centering
  \includegraphics[width=\textwidth]{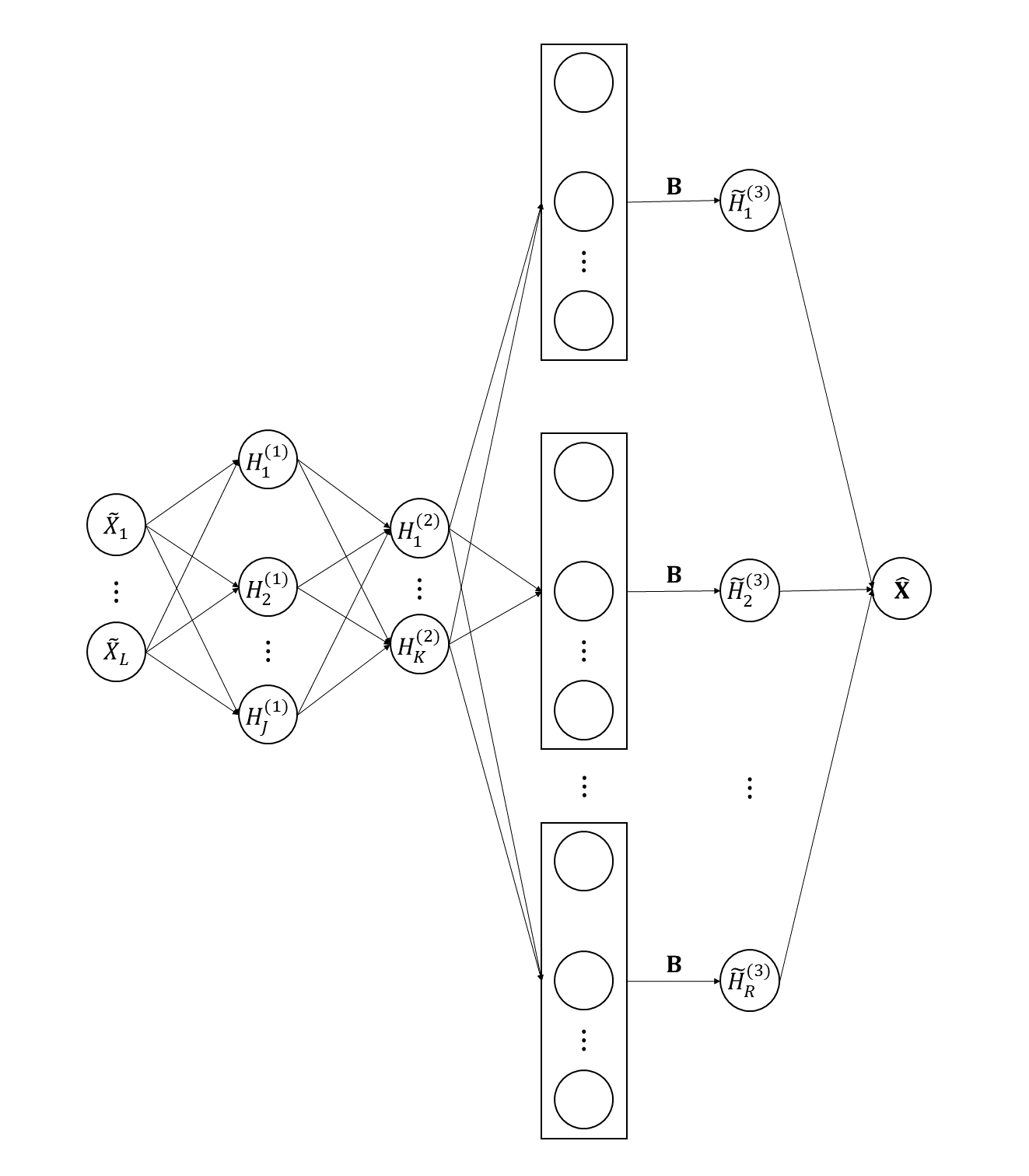}\\
  \caption{The transformed functional autoassociative neural network for nonlinear FPCA.}
  \label{FigNNtran}
\end{figure}

In practice, suppose that $X_1, \ldots, X_n$ are $n$ independent realizations of $X$, where $n$ is the sample size. Then, the loss function for the network can be represented as
\begin{align}
\frac{1}{n} \sum_{i = 1}^n \int_{\mathcal{T}} \{ X_i(t) - \wh{X}_i(t) \}^2 dt. \nonumber
\end{align}
However, integral appeared in the loss function can increase the difficulty of optimization. Hence, right-hand Riemann sum is employed for the computation. Specifically, let $0 = t_1 < t_2 < \ldots < t_M = 1$ be some equally spaced times points on $\mathcal{T}$. Moreover, denote $s_1, \ldots, s_T$ be the observation time points of the random curves, where $T$ is the observation size. Then, we consider the following loss function
\begin{align}
\wt{RE} = \frac{1}{n} \sum_{i = 1}^n \frac{1}{M - 1} \sum_{m = 2}^M \{ \wt{X}_i(t_m) - \wh{X}_i(t_m) \}^2, \label{REtilde}
\end{align}
where $\wt{X}_i(t_m)$ is the estimation of $X_i(t_m)$ using the observed data by smoothing.
Note that only values of the curves at $t_1, \ldots, t_M$ involved in the loss function. Therefore, we just need to consider discrete values of $X(t)$ in the last two layers of the network. Let $\textbf{B}_l = (B_l(t_0), \ldots, B_l(t_M))^\top$ for $l = 1, \ldots, L$, and $\textbf{B} = (\textbf{B}_1, \ldots, \textbf{B}_L)^\top$. The $r$th neuron of the third hidden layer can be computed by
\begin{align}
\wt{H}_r^{(3)} =  \sigma \Big \{ \Big ( A_r + \sum_{k = 1}^K V_{kr} H_k^{(2)} \Big )^\top \textbf{B} \Big \}. \label{NNtranlayerthree2}
\end{align}
Then the output is given by
\begin{align}
\wh{\textbf{X}} = \sum_{r = 1}^R u_r \wt{H}_r^{(3)}, \label{NNtranlayerfour}
\end{align}
where $\wh{\textbf{X}} = (\wh{X}(t_0), \ldots, \wh{X}(t_M))$. To be clear, the transformed functional autoassociative neural network is shown in Figure \ref{FigNNtran}. The computation involved in the transformed autoassociative neural network corresponds to (\ref{NNtranlayertwo}), (\ref{NNtranlayerthree2}) and (\ref{NNtranlayerfour}) respectively.

By turning the proposed functional autoassociative neural network in Section \ref{SecNN} into the transformed functional autoassociative neural network, the Adam algorithm \citep{kingma2014adam} can be employed in the optimization. This algorithm is popularly-used, and it can be realized by the Python package \texttt{torch}. Moreover, we also provide the Python package \texttt{nFunNN} for the implementation specific to our method.

\section{Simulation}\label{SecSim}

\subsection{Numerical performance}\label{SecSubNum}

In this section, we conduct a simulation study to explore the performance of the proposed nFunNN method. For the simulated data, we take into account the measurement error to make it more consistent with the practical cases. Specifically, the observed data is generated by
\begin{align}
Y_{ij} = X_i(s_j) + \epsilon_{ij}, \ i = 1, \ldots, n, \ j = 1, \ldots, T, \nonumber
\end{align}
where $Y_{ij}$ is the $j$th observation for the $i$th subject, and $\epsilon_{ij}$'s are the independent measurement errors, which we obtain from the normal distribution $\mathcal{N}(0, \delta^2)$. We set $\delta = 0.1$, and the observation size is set as $T = 51$. The observation grids are equally spaced on $\mathcal{T} = [0, 1]$. For the setting of $X_i$, we consider the following cases.
\begin{itemize}

\item Case 1: $X_i(t) = \xi_{i1} \sin (2 \pi t) + \xi_{i2} \cos (2 \pi t), \ t \in \mathcal{T}$, where $\xi_{i1}$'s and $\xi_{i2}$'s are simulated from $\mathcal{N} (0, 3^2)$ and $\mathcal{N} (0, 2^2)$, respectively.

\item Case 2: $X_i(t) = \xi_{i2} \sin (\xi_{i1} t), \ t \in \mathcal{T}$, where both $\xi_{i1}$'s and $\xi_{i2}$'s are simulated from $\mathcal{N} (0, 2^2)$.

\item Case 3: $X_i(t) = \xi_{i2} \cos (\xi_{i1} t), \ t \in \mathcal{T}$, where both $\xi_{i1}$'s and $\xi_{i2}$'s are simulated in the same way as that in Case 2.

\item Case 4: $X_i(t) = \xi_{i1} \sin (2 \pi t) + \xi_{i2} \cos (2 \pi t) + \xi_{i2} \sin (\xi_{i1} t), \ t \in \mathcal{T}$, where both $\xi_{i1}$'s and $\xi_{i2}$'s are simulated in the same way as that in Case 2.

\item Case 5: $X_i(t) = \xi_{i1} \sin (2 \pi t) + \xi_{i2} \cos (2 \pi t) + \xi_{i2} \cos (\xi_{i1} t), \ t \in \mathcal{T}$, where both $\xi_{i1}$'s and $\xi_{i2}$'s are simulated in the same way as that in Case 2.

\end{itemize}
The above setups include various structures of $X_i(t)$. In Case 1, $X_i(t)$ is actually generated through the Karhunen-Lo\`{e}ve expansion, with zero mean, $\lambda_1 = 3^2$, $\lambda_2 = 2^2$, and $\lambda_k = 0$ for $k \geq 3$. Thus, $X_i(t)$ in Case 1 has a linear structure, and a linear method may be suitable enough for this case. Moreover, the other four cases consider the nonlinear structure of $X_i(t)$. Case 2 and Case 3 impose only one nonlinear term in the setup, while Case 4 and Case 5 combine the linear terms in Case 1 with nonlinear terms in Case 2 and Case 3 respectively. Further, whether a linear structure or a nonlinear structure is considered, all the five cases are set to contain two principal components.

The proposed nFunNN method is compared with the classical linear FPCA method \citep{ramsay2005functional}. Specifically, the numbers of neurons in different layers for our transformed functional autoassociative network are set as $L = 10$, $J = 20$, $K = 2$, and $R = 20$. And the number of principal components for the linear FPCA method is selected as $2$.
To evaluate the performance of curve reconstruction, we consider the following criteria:
\begin{align}
\mbox{RMSE} &= \sqrt{\frac{1}{nM} \sum_{i = 1}^n \sum_{m = 1}^M \{ \wh{X}_i (t_m) - X_i (t_m) \}^2}, \nonumber \\
\mbox{RRMSE} &= \sqrt{\sum_{i = 1}^n \sum_{m = 1}^M \{ \wh{X}_i (t_m) - X_i (t_m) \}^2} \Bigg / \sqrt{\sum_{i = 1}^n \sum_{m = 1}^M X_i(t_m)^2 }, \nonumber
\end{align}
where $\mbox{RRMSE}$ means the relative $\mbox{RMSE}$.
Note that the prediction at $t_1, \ldots, t_M$ is assessed, and these time points can be different from the observation time points. In our simulation study, $t_1, \ldots, t_M$ are also equally spaced on $[0, 1]$ with $M$ being set as $101$. We consider the performance for both training set and test set in the evaluation, and the sizes of both sets are set as $1000$. Moreover, $100$ Monte Carlo runs are conducted for each considered case.

Table \ref{TabSim} lists the simulation results of the proposed nFunNN method and FPCA method in all the five cases. In Case 1, though FPCA method yields slightly better $\mbox{RMSE}$ and $\mbox{RRMSE}$ than our nFunNN method, both methods perform well for training set and test set. The result is not surprising, since $X_i(t)$ is generated by the Karhunen-Lo\`{e}ve expansion in Case 1, and classical linear FPCA is good enough to handle such case. For Case 2 and Case 3, where nonlinear structure is considered, it is evident that the proposed nFunNN method outperforms the FPCA method and gives more accurate prediction. That implies the advantage of our nFunNN method when solving nonlinear cases. Furthermore, FPCA method is almost invalid for Case 4 and Case 5, while the proposed nFunNN method still provides encouraging curve reconstruction results. In the setting of $X_i(t)$ in Case 4 and Case 5, we add a nonlinear term besides two linear terms, which makes the linear FPCA method cannot fulfill the prediction with only two principal components. It can be observed that the prediction error of our nFunNN method is much lower than that of the FPCA method in Case 4 and Case 5. That indicates our method can achieve more effective dimension reduction results in nonlinear settings.

To sum up, the proposed nFunNN method gives great predicting results for all the five cases. Although the error can be a bit larger than that of the linear method in linear case, the predicting results of our nFunNN method is still reasonable. Furthermore, the nFunNN method shows obvious superiority for the nonlinear cases. Therefore, our nFunNN method can be a good choice, when we have no idea whether the data at hand has a linear structure or a nonlinear structure.

\begin{table}[H]
\caption{The averaged $\mbox{RMSE}$ and $\mbox{RRMSE}$ of the nFunNN and FPCA methods across $100$ Monte Carlo runs for all the five cases, with standard deviation in parentheses.}
\label{TabSim}
\begin{center}
\begin{tabular}{cccccc}
\hline
 & & \multicolumn{2}{c}{Training set} & \multicolumn{2}{c}{Test set} \\
 & & \mbox{RASE} & \mbox{RRASE} & \mbox{RASE} & \mbox{RRASE} \\
\hline
\multirow{2}{*}{Case 1}& nFunNN &0.0285 (0.0088)&0.0112 (0.0034)&0.0324 (0.0187)&0.0127 (0.0072)\\
 &FPCA&0.0201 (0.0003)&0.0079 (0.0002)&0.0201 (0.0003)&0.0079 (0.0002)\\
\multirow{2}{*}{Case 2}& nFunNN &0.0453 (0.0146)&0.0386 (0.0121)&0.0607 (0.0182)&0.0519 (0.0154)\\
 &FPCA&0.0890 (0.0177)&0.0760 (0.0147)&0.0878 (0.0192)&0.0753 (0.0164)\\
\multirow{2}{*}{Case 3}& nFunNN &0.0790 (0.0250)&0.0488 (0.0156)&0.1036 (0.0300)&0.0639 (0.0182)\\
 &FPCA&0.1955 (0.0255)&0.1207 (0.0156)&0.2033 (0.0290)&0.1254 (0.0174)\\
\multirow{2}{*}{Case 4}& nFunNN &0.1828 (0.0576)&0.0791 (0.0252)&0.2181 (0.0746)&0.0943 (0.0320)\\
 &FPCA&0.9992 (0.0279)&0.4323 (0.0093)&1.0065 (0.0275)&0.4356 (0.0096)\\
\multirow{2}{*}{Case 5}& nFunNN &0.2248 (0.0514)&0.0877 (0.0199)&0.2781 (0.0637)&0.1080 (0.0244)\\
 &FPCA&0.6573 (0.0291)&0.2565 (0.0112)&0.6628 (0.0348)&0.2566 (0.0123)\\
\hline
\end{tabular}
\end{center}
\end{table}

\subsection{Effect of the tuning parameters}

In this section, we discuss the effect of $L$, $J$, $K$, and $R$ on the performance of the proposed nFunNN method. Though the values of $J$ and $R$ can be different for our method, we set $J = R$ for simplicity here. We consider the settings of Case 1, Case 2 and Case 4 for explanation in this section. For the discussion of the effect of $L$, we fix $J = 20$ and $K = 2$, and the value of $L$ can be selected by $10$, $15$, and $20$. Moreover, we set $J = 10, 15, 20$, and $L = 10$, $K = 2$ for the exploration of the influence of $J$. When discussing the effect of $K$, we set $K = 2, 3, 4$, $L = 10$ and $J = 20$. Note that the number of parameters in the network is $J ( KL + K + 2L + 2 )$ when $J = R$. As the sample size should be larger than the number of parameters, we increase the sample size of training set to $2000$, and the size of test set is still $1000$ as in Section \ref{SecSubNum}.

Tables \ref{TabSimL}--\ref{TabSimK} present the simulation results of our nFunNN method with the use of various $L$, $J$ and $K$. From Table \ref{TabSimL}, it can be observed that with the rise of $L$, both $\mbox{RMSE}$ and $\mbox{RRMSE}$ increase slightly. As shown in Table \ref{TabSimJ}, there is no obvious difference in the prediction errors with the use of various $J$, which implies that the effect of $J$ is not significant. For results in Table \ref{TabSimK}, the prediction errors for Case 1 are similar when different values of $K$ are considered. However, for Case 2 and Case 4, which are both nonlinear cases, various values of $K$ lead to different performance of the network. In Case 2, the prediction error first decreases with the increase of $K$, and then shows a minor growth. Furthermore, the performance of the nFunNN method gets much better when larger $K$ is used in Case 4. We conjecture the reason is related to the complex setting of Case 4.

To summarize, according to the simulation results, the effects of $L$ and $J$ are not very obvious, while different values of $K$ can bring large changes for the prediction in some complex cases. Moreover, the selection of tuning parameters in the neural network can be completed through the validation set.

\begin{table}[H]
\caption{The averaged $\mbox{RMSE}$ and $\mbox{RRMSE}$ of the nFunNN methods across $100$ Monte Carlo runs with the use of various values of $L$ for Case 1, Case 2 and Case 4, with standard deviation in parentheses.}
\label{TabSimL}
\begin{center}
\begin{tabular}{cccccc}
\hline
 & & \multicolumn{2}{c}{Training set} & \multicolumn{2}{c}{Test set} \\
 & & \mbox{RASE} & \mbox{RRASE} & \mbox{RASE} & \mbox{RRASE} \\
\hline
\multirow{3}{*}{Case 1}& $L = 10$ &0.0247 (0.0038)&0.0097 (0.0015)&0.0262 (0.0042)&0.0103 (0.0016)\\
 &$L = 15$&0.0301 (0.0060)&0.0119 (0.0024)&0.0312 (0.0062)&0.0122 (0.0024)\\
 &$L = 20$&0.0375 (0.0082)&0.0147 (0.0032)&0.0386 (0.0080)&0.0151 (0.0031)\\
\hline
\multirow{3}{*}{Case 2}& $L = 10$ &0.0414 (0.0084)&0.0354 (0.0071)&0.0522 (0.0162)&0.0443 (0.0138)\\
 &$L = 15$&0.0547 (0.0209)&0.0469 (0.0179)&0.0628 (0.0240)&0.0533 (0.0204)\\
 &$L = 20$&0.0708 (0.0309)&0.0605 (0.0261)&0.0780 (0.0318)&0.0663 (0.0276)\\
\hline
\multirow{3}{*}{Case 4}& $L = 10$ &0.1802 (0.0770)&0.0777 (0.0328)&0.1956 (0.0796)&0.0845 (0.0352)\\
 &$L = 15$&0.1881 (0.0678)&0.0812 (0.0293)&0.2018 (0.0763)&0.0870 (0.0327)\\
 &$L = 20$&0.2350 (0.1851)&0.1014 (0.0804)&0.2473 (0.1885)&0.1067 (0.0812)\\
\hline
\end{tabular}
\end{center}
\end{table}

\begin{table}[H]
\caption{The averaged $\mbox{RMSE}$ and $\mbox{RRMSE}$ of the nFunNN methods across $100$ Monte Carlo runs with the use of various values of $J$ for Case 1, Case 2 and Case 4, with standard deviation in parentheses.}
\label{TabSimJ}
\begin{center}
\begin{tabular}{cccccc}
\hline
 & & \multicolumn{2}{c}{Training set} & \multicolumn{2}{c}{Test set} \\
 & & \mbox{RASE} & \mbox{RRASE} & \mbox{RASE} & \mbox{RRASE} \\
\hline
\multirow{3}{*}{Case 1}& $J = 10$ &0.0228 (0.0031)&0.0090 (0.0012)&0.0251 (0.0060)&0.0099 (0.0023)\\
 &$J = 15$&0.0231 (0.0024)&0.0091 (0.0009)&0.0250 (0.0041)&0.0098 (0.0016)\\
 &$J = 20$&0.0246 (0.0035)&0.0097 (0.0014)&0.0262 (0.0043)&0.0103 (0.0017)\\
\hline
\multirow{3}{*}{Case 2}& $J = 10$ &0.0522 (0.0122)&0.0445 (0.0104)&0.0599 (0.0165)&0.0512 (0.0137)\\
 &$J = 15$&0.0457 (0.0112)&0.0390 (0.0094)&0.0541 (0.0147)&0.0463 (0.0125)\\
 &$J = 20$&0.0413 (0.0088)&0.0352 (0.0074)&0.0506 (0.0136)&0.0433 (0.0114)\\
\hline
\multirow{3}{*}{Case 4}& $J = 10$ &0.2102 (0.0507)&0.0907 (0.0220)&0.2266 (0.0612)&0.0978 (0.0259)\\
 &$J = 15$&0.1953 (0.0794)&0.0843 (0.0340)&0.2145 (0.0898)&0.0926 (0.0386)\\
 &$J = 20$&0.1684 (0.0413)&0.0727 (0.0179)&0.1854 (0.0507)&0.0801 (0.0224)\\
\hline
\end{tabular}
\end{center}
\end{table}

\begin{table}[H]
\caption{The averaged $\mbox{RMSE}$ and $\mbox{RRMSE}$ of the nFunNN methods across $100$ Monte Carlo runs with the use of various values of $K$ for Case 1, Case 2 and Case 4, with standard deviation in parentheses.}
\label{TabSimK}
\begin{center}
\begin{tabular}{cccccc}
\hline
 & & \multicolumn{2}{c}{Training set} & \multicolumn{2}{c}{Test set} \\
 & & \mbox{RASE} & \mbox{RRASE} & \mbox{RASE} & \mbox{RRASE} \\
\hline
\multirow{3}{*}{Case 1}& $K = 2$ &0.0248 (0.0042)&0.0097 (0.0016)&0.0258 (0.0045)&0.0102 (0.0018)\\
 &$K = 3$&0.0255 (0.0038)&0.0100 (0.0015)&0.0266 (0.0040)&0.0105 (0.0016)\\
 &$K = 4$&0.0263 (0.0029)&0.0103 (0.0012)&0.0275 (0.0036)&0.0108 (0.0015)\\
\hline
\multirow{3}{*}{Case 2}& $K = 2$ &0.0394 (0.0082)&0.0337 (0.0071)&0.0512 (0.0185)&0.0436 (0.0158)\\
 &$K = 3$&0.0271 (0.0016)&0.0232 (0.0015)&0.0349 (0.0133)&0.0297 (0.0114)\\
 &$K = 4$&0.0290 (0.0015)&0.0248 (0.0014)&0.0352 (0.0096)&0.0300 (0.0082)\\
\hline
\multirow{3}{*}{Case 4}& $K = 2$ &0.1733 (0.0522)&0.0748 (0.0224)&0.1899 (0.0663)&0.0820 (0.0283)\\
 &$K = 3$&0.0610 (0.0084)&0.0263 (0.0036)&0.0754 (0.0158)&0.0326 (0.0067)\\
 &$K = 4$&0.0334 (0.0027)&0.0144 (0.0012)&0.0405 (0.0079)&0.0175 (0.0033)\\
\hline
\end{tabular}
\end{center}
\end{table}

\section{Real Data Analysis}\label{SecReal}

In this section, we discuss the performance of the proposed nFunNN method for real data application. Yoga data set and StarLightCurves data set from \citep{UCRArchive} are considered. In specific, we aim to assess the predicting ability of the nFunNN method by these two data sets.

\begin{figure}
  \centering
  \includegraphics[width=0.9\textwidth]{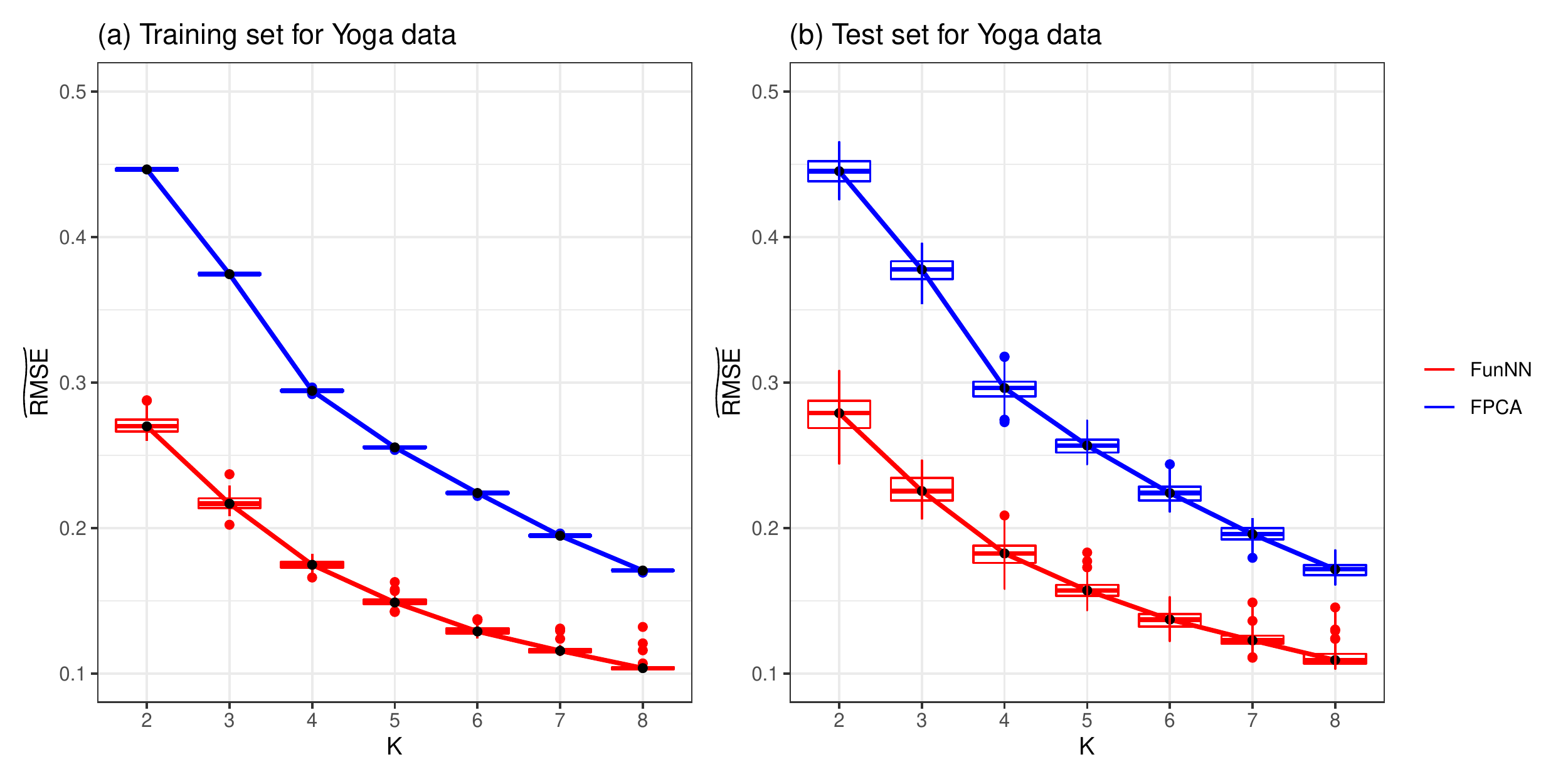}\\
  \caption{The averaged $\wt{\mbox{RMSE}}$ of the nFunNN and FPCA methods across $100$ runs for the Yoga data set. (a) The boxplot for training set. (b) The boxplot for test set.}
  \label{FigRealYoga}
\end{figure}

For the Yoga data set, it contains $3300$ samples and the observation size is $426$ for each subject. We randomly divide the data set into training set and test set with the sample size being $3000$ and $300$ respectively. The values at these $426$ observation grids are predicted by both our nFunNN method and the classical FPCA method via different $K$. The numbers of neurons in other layers for our transformed functional autoassociative neural network are set as $L = 20$, $J = 20$, and $R = 20$. The following criteria are considered:
\begin{align}
\wt{\mbox{RMSE}} &= \sqrt{\frac{1}{nM} \sum_{i = 1}^n \sum_{m = 1}^M \{ \wh{X}_i (t_m) - Y_{im} \}^2}, \nonumber \\
\wt{\mbox{RRMSE}} &= \sqrt{\sum_{i = 1}^n \sum_{m = 1}^M \{ \wh{X}_i (t_m) - Y_{im} \}^2} \Bigg / \sqrt{\sum_{i = 1}^n \sum_{m = 1}^M Y_{im}^2 }, \nonumber
\end{align}
where $M = 426$ for the Yoga data set. The data set is randomly split for $100$ times, and the predicting results for both training set and test set are presented in Table \ref{TabRealYoga}. It is shown that with the rise of $K$, the predicting results are getting better for both nFunNN and FPCA methods. Moreover, the proposed nFunNN method can provide more precise predicting results when using the same $K$ as the FPCA method, which demonstrates the advantage of our method in real data application. Figure \ref{FigRealYoga} exhibits the predicting performance of both methods for training set and test set by boxplots. It can be observed that nFunNN method always produces less predicting error.

\begin{table}[htbp]
\caption{The averaged $\wt{\mbox{RMSE}}$ and $\wt{\mbox{RRMSE}}$ of the nFunNN and FPCA methods across $100$ runs for the Yoga data set, with standard deviation in parentheses.}
\label{TabRealYoga}
\begin{center}
\begin{tabular}{cccccc}
\hline
 & & \multicolumn{2}{c}{Training set} & \multicolumn{2}{c}{Test set} \\
 & & $\wt{\mbox{RMSE}}$ & $\wt{\mbox{RMMSE}}$ & $\wt{\mbox{RMSE}}$ & $\wt{\mbox{RMMSE}}$ \\
\hline
\multirow{2}{*}{$K = 2$}& FunNN &0.2709 (0.0061)&0.2712 (0.0061)&0.2777 (0.0127)&0.2780 (0.0127)\\
 &FPCA&0.4464 (0.0009)&0.4469 (0.0009)&0.4453 (0.0093)&0.4459 (0.0093)\\
\multirow{2}{*}{$K = 3$}& FunNN &0.2172 (0.0052)&0.2174 (0.0052)&0.2265 (0.0097)&0.2267 (0.0097)\\
 &FPCA&0.3745 (0.0009)&0.3750 (0.0009)&0.3769 (0.0085)&0.3773 (0.0085)\\
\multirow{2}{*}{$K = 4$}& FunNN &0.1747 (0.0032)&0.1749 (0.0032)&0.1821 (0.0089)&0.1823 (0.0089)\\
 &FPCA&0.2943 (0.0009)&0.2947 (0.0009)&0.2952 (0.0092)&0.2955 (0.0092)\\
\multirow{2}{*}{$K = 5$}& FunNN &0.1492 (0.0032)&0.1494 (0.0032)&0.1572 (0.0068)&0.1574 (0.0068)\\
 &FPCA&0.2554 (0.0007)&0.2557 (0.0007)&0.2565 (0.0068)&0.2568 (0.0068)\\
\multirow{2}{*}{$K = 6$}& FunNN &0.1294 (0.0029)&0.1295 (0.0029)&0.1366 (0.0059)&0.1367 (0.0059)\\
 &FPCA&0.2240 (0.0006)&0.2243 (0.0006)&0.2242 (0.0063)&0.2245 (0.0063)\\
\multirow{2}{*}{$K = 7$}& FunNN &0.1159 (0.0028)&0.1161 (0.0028)&0.1232 (0.0055)&0.1234 (0.0055)\\
 &FPCA&0.1947 (0.0005)&0.1949 (0.0005)&0.1955 (0.0055)&0.1958 (0.0055)\\
\multirow{2}{*}{$K = 8$}& FunNN &0.1043 (0.0036)&0.1044 (0.0036)&0.1110 (0.0063)&0.1111 (0.0063)\\
 &FPCA&0.1707 (0.0005)&0.1709 (0.0005)&0.1714 (0.0049)&0.1716 (0.0049)\\
\hline
\end{tabular}
\end{center}
\end{table}

\begin{figure}
  \centering
  \includegraphics[width=0.9\textwidth]{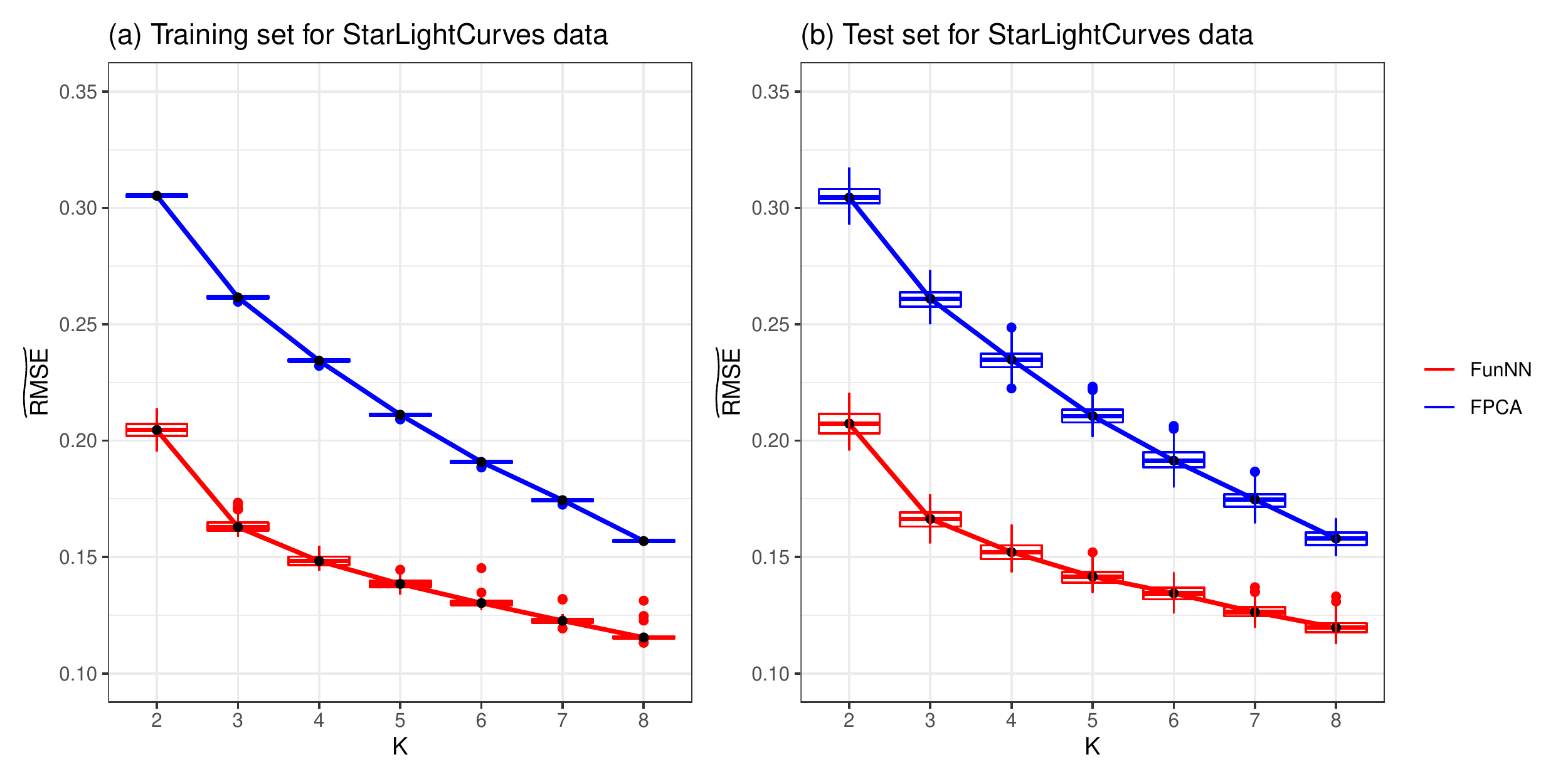}\\
  \caption{The averaged $\wt{\mbox{RMSE}}$ of the nFunNN and FPCA methods across $100$ runs for the StarLightCurves data set. (a) The boxplot for training set. (b) The boxplot for test set.}
  \label{FigRealStar}
\end{figure}

Furthermore, we consider the analysis of the StarLightCurves data set. There are $9236$ subjects in this data set and each subject has $1024$ observations. Similar to the analysis of Yoga data set, we randomly divide the StarLightCurves data set into training set and test set, and the sizes are set as $8000$ and $1236$ respectively. We intend to predict the values at these $1024$ observation grids by both nFunNN and FPCA methods using various $K$. The numbers of the neurons for the transformed functional autoassociative neural network are set as the same as that in the analysis of Yoga data set. We also conduct $100$ runs for StarLightCurves data set, and the averaged $\wt{\mbox{RMSE}}$ and $\wt{\mbox{RRMSE}}$ are reported in Table \ref{TabRealStar}. The trend of the prediction error for both methods is analogous with that for Yoga data set. Figure \ref{FigRealStar} provides visual illustration for the comparison of nFunNN and FPCA methods. It is shown that our nFunNN method constantly outperforms the linear FPCA method with the use of different $K$. That further indicates the effectiveness of the proposed nFunNN method in real-world application.

\begin{table}[htbp]
\caption{The averaged $\wt{\mbox{RMSE}}$ and $\wt{\mbox{RRMSE}}$ of the nFunNN and FPCA methods across $100$ runs for the StarLightCurves data set, with standard deviation in parentheses.}
\label{TabRealStar}
\begin{center}
\begin{tabular}{cccccc}
\hline
 & & \multicolumn{2}{c}{Training set} & \multicolumn{2}{c}{Test set} \\
 & & $\wt{\mbox{RMSE}}$ & $\wt{\mbox{RMMSE}}$ & $\wt{\mbox{RMSE}}$ & $\wt{\mbox{RMMSE}}$ \\
\hline
\multirow{2}{*}{$K = 2$}& FunNN &0.2048 (0.0039)&0.2049 (0.0039)&0.2074 (0.0057)&0.2075 (0.0057)\\
 &FPCA&0.3052 (0.0007)&0.3053 (0.0007)&0.3047 (0.0046)&0.3048 (0.0046)\\
\multirow{2}{*}{$K = 3$}& FunNN &0.1636 (0.0032)&0.1637 (0.0032)&0.1662 (0.0045)&0.1663 (0.0045)\\
 &FPCA&0.2615 (0.0007)&0.2617 (0.0007)&0.2609 (0.0047)&0.2611 (0.0047)\\
\multirow{2}{*}{$K = 4$}& FunNN &0.1484 (0.0023)&0.1485 (0.0023)&0.1519 (0.0044)&0.1520 (0.0044)\\
 &FPCA&0.2344 (0.0008)&0.2345 (0.0008)&0.2346 (0.0050)&0.2347 (0.0050)\\
\multirow{2}{*}{$K = 5$}& FunNN &0.1386 (0.0019)&0.1387 (0.0019)&0.1415 (0.0033)&0.1416 (0.0033)\\
 &FPCA&0.2110 (0.0007)&0.2112 (0.0007)&0.2110 (0.0044)&0.2111 (0.0044)\\
\multirow{2}{*}{$K = 6$}& FunNN &0.1305 (0.0020)&0.1305 (0.0020)&0.1344 (0.0032)&0.1344 (0.0032)\\
 &FPCA&0.1908 (0.0007)&0.1909 (0.0007)&0.1919 (0.0048)&0.1920 (0.0048)\\
\multirow{2}{*}{$K = 7$}& FunNN &0.1228 (0.0017)&0.1229 (0.0017)&0.1267 (0.0032)&0.1267 (0.0032)\\
 &FPCA&0.1744 (0.0007)&0.1745 (0.0007)&0.1747 (0.0044)&0.1748 (0.0044)\\
\multirow{2}{*}{$K = 8$}& FunNN &0.1158 (0.0021)&0.1158 (0.0021)&0.1198 (0.0031)&0.1199 (0.0031)\\
 &FPCA&0.1568 (0.0006)&0.1569 (0.0006)&0.1580 (0.0037)&0.1581 (0.0037)\\
\hline
\end{tabular}
\end{center}
\end{table}

\section{Conclusions and Discussion}\label{SecDis}

In this paper, we introduce a nonlinear FPCA method to realize effective dimension reduction and curve reconstruction. We generalize the autoassociative neural network to our functional data analysis framework and construct a transformed functional autoassociative neural network for practical implementation. The proposed method takes into account the nonlinear structure of the functional observations. A Python package is developed for the convenience of using the proposed nFunNN method. The theoretical properties of the proposed networks are also considered. Moreover, the results of the simulation study and real data application further suggest the superiority of our nFunNN method.

There are also several possible extension for our work. First, we only consider usual functional data in the development of our method. However, complex function data, such as multivariate functional data \citep{chiou2014multivariate} and multidimensional functional data \citep{wang2022low}, becomes more and more common nowadays. Thus, considering nonlinear FPCA method for these types of data and generalizing our method to solve such issue can be of great significance. Second, only curve reconstruction error is considered in the construction of the loss function (\ref{REtilde}) for our method. So, our method is particularly suitable for the curve reconstruction issue. It can be beneficial if other concerns can be imposed in the loss function, such as regression and clustering problems. To achieve this goal, some modifications of the proposed neural network are needed, which is worth further research.

\section*{Acknowledgments}

This work was supported by Public Health $\&$ Disease Control and Prevention, Major Innovation $\&$ Planning Interdisciplinary Platform for the ``Double-First Class" Initiative, Renmin University of China. This work was also supported by the Outstanding Innovative Talents Cultivation Funded Programs 2021 of Renmin University of China.

\appendix
\section{ Proof of Theorem \ref{TheUA} }

\begin{proof}

Let $\mathbb{K}$ be a compact subset of $L^2(\mathcal{T})$.
Recall that $\mathcal{S} (\sigma, L^2(\mathcal{T}))$ is the set of functions from $L^2(\mathcal{T})$ to $\mathbb{R}$ of the form
\begin{align}
X \mapsto \sum_{j = 1}^J w_{j0} \sigma \Big \{ b_j + \int_{\mathcal{T}} X(t) \beta_j(t) dt \Big \}, \nonumber
\end{align}
where $J \in \mathbb{N}^{\ast}$, $w_{j0} \in \mathbb{R}$, $b_j \in \mathbb{R}$ and $\beta_j(t) \in L^2(\mathcal{T})$. According to Corollary 5.1.2 in \citep{stinchcombe1999neural} and Section 6.1.2 of \citep{rossi2006theoretical}, $\mathcal{S} (\sigma, L^2(\mathcal{T}))$ has the universal approximation property. Hence, for any continuous function $F$ from $\mathbb{K}$ to $\mathbb{R}$ and for any $\epsilon > 0$, there exist $H \in \mathcal{S} (\sigma, L^2(\mathcal{T}))$ such that for all $X \in \mathbb{K}$,
\begin{align}
|H(X) - F(X)| < \epsilon / 2. \label{theoryHF}
\end{align}

As $H$ is continuous in $L^2(\mathcal{T})$, we have that for any $X \in \mathbb{K}$, there exist $\eta(X) > 0$ such that for any $f \in B(X, \eta(X))$, $|H(X) - H(f)| < \epsilon / 4$. By the approximation property of B-splines \citep{zhong2023sparse} and the compactness of $\mathbb{K}$, we can get
\begin{align}
| H(\wt{X}) - H(X)| < \epsilon / 2, \label{theoryHH}
\end{align}
similar to the proof in Section 6.1.3 of \citep{rossi2006theoretical}, where $\wt{X}(t) = \sum_{h = 1}^L x_h B_h(t)$. Then according to (\ref{theoryHF}) and (\ref{theoryHH}),
\begin{align}
| H(\wt{X}) - F(X)| \leq | H(\wt{X}) - H(X)| + |H(X) - F(X)| < \epsilon. \nonumber
\end{align}
Moreover,
\begin{align}
H(\wt{X}) &= \sum_{j = 1}^J w_{j0} \sigma \Big \{ b_j + \int_{\mathcal{T}} \wt{X}(t) \beta_j(t) dt \Big \} \nonumber \\
&= \sum_{j = 1}^J w_{j0} \sigma \Big [ b_j + \int_{\mathcal{T}} \Big \{ \sum_{h = 1}^L x_h B_h(t) \Big \} \beta_j(t) dt \Big ] \nonumber \\
&= \sum_{j = 1}^J w_{j0} \sigma \Big \{ b_j + \sum_{h = 1}^L x_h \int_{\mathcal{T}} \beta_j(t) B_h(t) dt \Big \} \nonumber \\
&= \sum_{j = 1}^J w_{j0} \sigma \Big ( b_j + \sum_{k = 1}^{L} d_{jh} x_h \Big ). \nonumber
\end{align}
Define $\wt{H}(X) = H(\wt{X})$, then we have $\wt{H} \in \mathcal{S} (\sigma, \textbf{B}_L)$. The proof is completed.

\end{proof}

\bibliographystyle{unsrtnat}
\bibliography{ref}

\end{document}